\documentclass[pra,groupedaddress,showpacs,floatfix]{revtex4}

\usepackage{amssymb,amsmath}
\usepackage{graphics}

\begin{document}

\title{An introduction to quantum game theory}
\author{A.P.\ Flitney}
\email{aflitney@physics.adelaide.edu.au}
\author{D.\ Abbott}
\email{dabbott@eleceng.adelaide.edu.au}
\affiliation{Centre for Biomedical Engineering (CBME)
and Department of Electrical and Electronic Engineering, \\
The University of Adelaide , SA 5005, Australia}
\date{11 August 2002}

\begin{abstract}
The application of the methods of quantum mechanics
to game theory provides us with the ability to achieve results
not otherwise possible.
Both linear superpositions of actions
and entanglement between the players' moves can be exploited.
We provide an introduction to quantum game theory
and review the current status of the subject. 
\end{abstract}


\maketitle

\section{Introduction}
Game theory is the study of decision making of competing agents
in some conflict situation.
It was developed in the 1930's
in the context of economic theory
and was first formalized by von Neumann and Morgenstern
in {\em The Theory of Games and Economic Behavior}~\cite{neumann44}.
Subsequently much work was done on game theory
and it is now a mature discipline used in areas
such as the social sciences, biology and engineering.

With the recent interest in quantum computing
and quantum information theory,
there has been an effort to recast classical game theory
using quantum probability amplitudes,
and hence study the effect of quantum superposition,
interference and entanglement on the agents' optimal strategies.
Apart from unsolved problems in quantum information theory~\cite{god},
quantum game theory may be useful in studying quantum communication
since that can be considered as a game
where the objective is to maximize effective communication.
Inspired by quantum game theory, within the new branch of mathematical
economics, econophysics, workers have attempted to model markets and auctions,
assuming traders can use quantum protocols~\cite{piotr01a,piotr01b}.
Here, a quantum strategy, representing a superposition of trading decisions,
can give an advantage over classical strategies~\cite{piotr02}.
This, of course, must be distinguished from attempts to use the
mathematical machinery of quantum field theory to solve {\em classical}
financial market problems~\cite{ilinski,baaquie01}.

There has been recent work linking quantum games
to the production of algorithms for quantum computers~\cite{lee02}
that may lead to new ways of approaching such algorithms.
There is also speculation that nature may be playing quantum games
at the molecular level~\cite{eisert99}.
In support of this, recent work on protein molecules,
including the protein folding problem, is now turning to a full quantum
mechanical description at the molecular level~\cite{gogonea99}.

The seminal work on quantum game theory by Meyer~\cite{meyer99}
studied a simple coin tossing game
and showed how a player utilizing quantum superposition
could win with certainty against a classical player.
A general protocol for two player--two strategy quantum games
with entanglement was developed by Eisert {\em et al}~\cite{eisert99}
using the well known prisoners' dilemma as an example
and this was extended to multiplayer games
by Benjamin and Hayden~\cite{benjamin00b}.

Since then more work has been done on quantum prisoners'
dilemma~\cite{benjamin00a,eisert00,du01a,du01b,du01c,du02,iqbal01a,iqbal02}
and a number of other games have been converted
to the quantum realm including
the battle of the sexes~\cite{iqbal01a,marinatto00,du00a,iqbal01},
the Monty Hall problem~\cite{li01,flitney02a,dariano02},
rock-scissors-paper~\cite{iqbal01b}, and others
\cite{du00b,benjamin00b,kay01,iqbal01c,iqbal02a,iqbal02b,iqbal02c,johnson01,guinea02}.

In section two we introduce some terminology
and basic ideas of game theory.
Section three will introduce the idea of quantum game theory
and give the general protocol for the construction of such games,
along with a review of the existing work in this field.
Finally, in section four we
discuss some questions and suggestions.

\section{Game Theory}
\subsection{Basic ideas and terminology}
Game theory attempts to mathematically model a situation
where agents interact.
The agents in the game are called {\em players},
their possible actions {\em moves},
and a prescription that specifies the particular move
to be made in all possible game situations
a {\em strategy}.
That is, a strategy represents a plan of action
that contains all the contingencies that can possibly arise
within the rules of the game.
In response to some particular game situation,
a {\em pure} strategy consists of always playing a
given move,
while a strategy that utilizes a randomizing device to select between
different moves is known as a {\em mixed} strategy.
The {\em utility} to a player of a game outcome
is a numerical measure of the desirability of that outcome
for the player.
A {\em payoff matrix} gives numerical values to the players' utility
for all the game outcomes.
It is assumed that the players will seek to maximize their utility within
the given rules of the game.
Games in which the choices of the players are known as soon as they are made
are called games of {\em perfect information}.
These are the main ones that are of interest to us here.

A {\em dominant} strategy
is one that does at least as well as any competing strategy
against any possible moves by the other player(s).
The {\em Nash equilibrium} (NE) is the most important of the possible
equilibria in game theory.
It is the combination of strategies from which no player
can improve his/her payoff by a unilateral change of strategy.
A {\em Pareto optimal} outcome is one
from which no player can obtain a higher utility
without reducing the utility of another.
Strategy $A$ is {\em evolutionary stable} against $B$ if,
for all sufficiently small, positive $\epsilon$,
$A$ performs better than $B$ against the mixed strategy
$(1-\epsilon) A + \epsilon B$.
An {\em evolutionary stable strategy} (ESS)~\cite{msmith73}
is one that is evolutionary stable against all other strategies.
The set of all strategies that are ESS is a subset of the NE of the game.
A two player {\em zero-sum} game is one
where the interests of the players are diametrically opposed.
That is, the sum of the payoffs for any game result is zero.
In such a game a {\em saddle point} is an entry in the payoff matrix
for (say) the row player that is both the minimum of its row
and the maximum of its column.

\subsection{An example: the prisoners' dilemma}
A two player game where each player has two possible moves
is known as a $2 \times 2$ game,
with obvious generalizations to larger strategic spaces
or number of players.
As an example,
consider one such game that has deservedly received much attention:
the prisoners' dilemma.
Here the players' moves are known as
cooperation ($C$) or defection ($D$).
The payoff matrix is such that
there is a conflict between the NE
and the Pareto optimal outcome.
The payoff matrix can be written as
\begin{equation}
\label{e-pd}
  \begin{array}{c|cc}
	 & \mbox{Bob}: C & \mbox{Bob}: D \\
	\hline
	\mbox{Alice}: C & (3,3) & (0,5) \\
	\mbox{Alice}: D & (5,0) & (1,1)
  \end{array}
\end{equation}
where the numbers in parentheses represent the row (Alice)
and column (Bob) player's payoffs, respectively.
The game is symmetric and there is a dominant strategy,
that of always defecting,
since it gives a better payoff if the other player cooperates
(five instead of three)
or if the other player defects
(one instead of zero).
Where both players have a dominant strategy
this combination is the NE.

The NE outcome $\{D,D\}$ is not such a good one
for the players, however,
since if they had both cooperated
they would have both received a payoff of three,
the Pareto optimal result.
In the absence of communication or negotiation
we have a dilemma,
some form of which is responsible for much of the misery and conflict
through out the world.

\section{Quantum Game Theory}
\subsection{Introductory ideas: penny flip}
A two state system, such as a coin,
is one of the simplest gaming devices.
If we have a player that can utilize quantum moves
we can demonstrate how the expanded space of possible strategies can
be turned to advantage.
Meyer, in his seminal work on quantum game theory~\cite{meyer99},
considered the simple game ``penny flip''
that consists of the following:
Alice prepares a coin in the heads state,
Bob, without knowing the state of the coin,
can choose to either flip the coin or leave its state unaltered,
and Alice, without knowing Bob's action,
can do likewise.
Finally, Bob has a second turn at the coin.
The coin is now examined and Bob wins if it shows heads.
A classical coin clearly gives both players an equal probability of success
unless they utilize knowledge of the other's psychological bias,
and such knowledge is beyond analysis by standard game theory~\footnote{
The biases of opposing players can be modeled with game theory but additional
formalism is required~\cite{rubinstein98}.}.

To quantize this game,
we replace the coin by a two state quantum system
such as a spin one-half particle.
Now Bob is given the power to make quantum moves while Alice is
restricted to classical ones.
Can Bob profit from his increased strategic space?
Let $|0\rangle$ represent the ``heads'' state
and $|1\rangle$ the ``tails'' state.
Alice initially prepares the system in the $|0\rangle$ state.
Bob can proceed by first applying the Hadamard operator,
\begin{equation}
\hat{H} = \frac{1}{\sqrt{2}} \left(\begin{array}{cc}
		1 & 1 \\
		1 & -1
	  \end{array}	\right) \;,
\end{equation}
putting the system into the equal superposition of the two states:
$1/\sqrt{2} (|0\rangle + |1\rangle)$.
Now Alice can leave the ``coin'' alone
or interchange the states $|0\rangle$ and $|1\rangle$,
but if we suppose this is done without causing the system to decohere
either action will leave the system unaltered,
a fact that can be exploited by Bob.
In his second move he applies the Hadamard operator again
resulting in the pure state $|0\rangle$
thus winning the game.
Bob utilized a superposition of states
and the increased latitude allowed him by the possibility
of quantum operators to make Alice's strategy irrelevant,
giving him a certainty of winning.

We shall see later that quantum enhancement often exploits entangled states,
but in this case it is just the increased possibilities available to the
quantum player that proved decisive.
Du {\em et al} has also considered quantum strategies
in a simplified card game that do not rely
on entanglement~\cite{du00b}.

\subsection{A general prescription}
Where a player has a choice of two moves
they can be encoded by a single bit.
To translate this into the quantum realm
we replace the bit by a quantum bit or qubit
that can be in a linear superposition of the two states.
The basis states $|0\rangle$ and $|1\rangle$
correspond to the classical moves.
The players' qubits are initially prepared in some state
to be specified later.
We suppose that the players have a set of instruments that can manipulate
their qubit to apply their strategy
without causing decoherence of the quantum state.
That is, a pure quantum strategy is a unitary operator acting on the player's
qubit.
Unitary operations on the pair of qubits
can be carried out either before the players' moves,
for example to entangle the qubits,
or afterwards, for example,
to disentangle them
or to chose an appropriate basis for measurement.
Finally, a measurement
in the computational basis $\{|0\rangle, |1\rangle \}$
is made on the resulting state
and the payoffs are determined in accordance with the payoff matrix.
Knowing the final state prior to the measurement,
the expectation values of the payoffs can be calculated.
The identity operator $\hat{I}$ corresponds to retaining the initial choice while
\begin{equation}
\hat{F} \equiv i \hat{\sigma}_x = \left( \begin{array}{cc}
					0 & i \\
					i & 0
				  \end{array} \right)
\end{equation}
corresponds to a bit flip.
The resulting quantum game should contain the classical one as a subset.
 
We can extend the list of possible quantum actions
to include any physically realizable action on a player's qubit
that is permitted by quantum mechanics.
Some of the actions that have been considered
include projective measurement and entanglement
with ancillary bits or qubits.

A quantum game of the above form is easily realized as a quantum algorithm.
Physical simulation of such an algorithm has already been performed
for a quantum prisoners' dilemma in a two qubit nuclear magnetic resonance
computer~\cite{du01a}.

\subsection{Quantum $2 \times 2$ games}
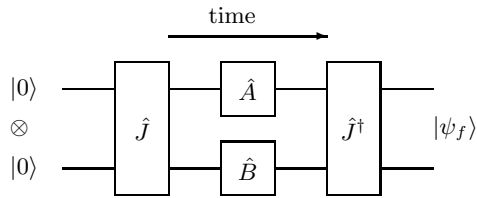
\begin{figure}
\begin{picture}(180,70)(-80,0)
	\multiput(0,8)(0,30){2}{$|0\rangle$}
	\put(0,23){$\otimes$}
	\put(160,23){$|\psi_f\rangle$}

	\multiput(20,10)(40,0){4}{\line(1,0){20}}
	\multiput(20,40)(40,0){4}{\line(1,0){20}}
	\put(40,0){\framebox(20,50){$\hat{J}$}}
	\put(120,0){\framebox(20,50){$\hat{J}^{\dagger}$}}

	\put(80,30){\framebox(20,20){$\hat{A}$}}
	\put(80,0){\framebox(20,20){$\hat{B}$}}

	\put(60,60){\vector(1,0){60}}
	\put(75,65){time}
\end{picture}
\caption{A general protocol for a two person quantum game
showing the flow of information.
$\hat{A}$ is Alice's move, $\hat{B}$ is Bob's,
and $\hat{J}$ is an entangling gate.}
\label{f-game22}
\end{figure}

In traditional $2 \times 2$ games
where each player has just a single move,
creating a superposition by utilizing a quantum strategy
will give the same results
as a mixed classical strategy.
In order to see non-classical results
Eisert {\em et al}~\cite{eisert99}
produced entanglement between the players' moves.
Keeping in mind that the classical game is to be a subset of the quantum one,
Eisert created the protocol in Fig.~\ref{f-game22} for a quantum game
between two players, Alice and Bob.

The final state can be computed by
\begin{equation}
| \psi_f \rangle = \hat{J}^{\dagger} (\hat{A} \otimes \hat{B}) \hat{J}
			| \psi_i \rangle \;,
\end{equation}
where
$|\psi_i \rangle = |00\rangle$ represents the initial state of the qubits and
$|\psi_f \rangle$ the final states,
$\hat{J}$ is an operator that entangles the players' qubits,
and $\hat{A}$ and $\hat{B}$ represent Alice's and Bob's move, respectively.
A disentangling gate $\hat{J^{\dagger}}$ is applied
prior to taking a measurement on the final state 
and the payoff is subsequently computed from the
classical payoff matrix.
Since we require the classical game to be a subset of the quantum one,
of necessity $\hat{J}$ commutes
with the direct product of any pair of classical moves.
In the quantum game it is only
the expectation value of the players' payoffs that is important.
For Alice (Bob) we can write
\begin{equation}
\langle \$ \rangle = P_{00} |\langle \psi_f|00 \rangle|^2 +
			P_{01} |\langle \psi_f|01 \rangle|^2 +
			P_{10} |\langle \psi_f|10 \rangle|^2 +
			P_{11} |\langle \psi_f|11 \rangle|^2
\end{equation}
where $P_{ij}$ is the payoff for Alice (Bob) associated with the game outcome
$ij, \; i, j \in \{0,1\}$.
If both players apply classical strategies
the quantum game provides nothing new.
However, if the players adopt quantum strategies
the entanglement provides the opportunity for the players'
moves to  interact in ways with no classical analogue.

A maximally entangling operator $\hat{J}$,
for an $N \times 2$ game, may be written,
without loss of generality~\cite{benjamin00b}, as
\begin{equation}
\label{e-entangle}
\hat{J} = \frac{1}{\sqrt{2}} (\hat{I}^{\otimes N}
		+ i \hat{\sigma}_{x}^{\otimes N}) \;.
\end{equation}
An equivalent form of the entangling operator
that permits the degree of entanglement to be controlled by a parameter
$\gamma \in [0, \pi/2]$ is
\begin{equation}
\label{e-entangle2}
\hat{J} = \exp \left( i \frac{\gamma}{2} \hat{\sigma}_{x}^{\otimes N} \right) \;,
\end{equation}
with maximal entanglement corresponding to $\gamma = \pi/2$.

The full range of pure quantum strategies
are any $\hat{U} \in$ SU(2).
We may write
\begin{equation}
\label{e-su2}
\hat{U}(\theta, \alpha, \beta) =
	\left( \begin{array}{cc}
		e^{i \alpha} \cos (\theta/2) & i e^{i \beta} \sin (\theta/2) \\
		i e^{-i \beta} \sin (\theta/2) & e^{-i \alpha} \cos(\theta/2)
	    \end{array} \right) \;,
\end{equation} 
where $\theta \in [0,\pi]$ and $\alpha, \beta \in [-\pi,\pi]$.
The strategies $\tilde{U}(\theta) \equiv \hat{U}(\theta, 0, 0)$
are equivalent to classical mixtures between
the identity and bit flip operations.
When Alice plays $\tilde{U}(\theta_A)$ and Bob plays $\tilde{U}(\theta_B)$
the payoffs are separable functions of $\theta_A$ and $\theta_B$
and we have nothing more than could be obtained from the classical game
by employing mixed strategies.

In quantum prisoners' dilemma a player with access to quantum strategies can always do
at least as well as a classical player.
If cooperation is associated with the $|0\rangle$ state
and defection with the $|1\rangle$ state,
then the strategy ``always cooperate'' is $\hat{C} \equiv \tilde{U}(0) = \hat{I}$
and the strategy ``always defect'' is $\hat{D} \equiv \tilde{U}(\pi) = \hat{F}$.
Against a classical Alice playing $\tilde{U}(\theta)$,
a quantum Bob can play Eisert's ``miracle'' move~\footnote{There are some
notational differences to Eisert {\em et al}'s original paper~\cite{eisert99}
in the form of $\hat{D}$ that
necessitates a corresponding change in $\hat{J}$.
This allows an easier generalization of the entanglement operator
to multiplayer games.
The only effect on the results is a possible rotation of $|\psi_f\rangle$
in the complex plane that is not physically observable.}
\begin{equation}
\hat{M} = \hat{U}(\frac{\pi}{2}, \frac{\pi}{2}, 0) = \frac{i}{\sqrt{2}}
	\left(  \begin{array}{cc}
			1 & 1 \\
			1 & -1
		\end{array} \right)
\end{equation}
that yields a payoff of $\langle \$_B \rangle = 3 + 2 \sin \theta$ for Bob
while leaving Alice with only $\langle \$_A \rangle = 1/2 \, (1 - \sin \theta)$.
In this case the dilemma is removed in favor of the quantum player.
In the partially entangled case,
there is a critical value of the entanglement parameter
$\gamma = \arcsin (1/\sqrt{5})$,
below which the quantum player should revert
to the classical dominant strategy $\hat{D}$
to ensure a maximal payoff~\cite{eisert99}.
At the critical level of entanglement there is effectively a phase change between
the quantum and classical domains of the game~\cite{du01c,du02}.

In a space of restricted quantum strategies,
corresponding to setting $\beta=0$ in Eq.~(\ref{e-su2}),
Eisert demonstrated that there was a new NE
that yielded a payoff of three to both players,
the same as mutual cooperation.
This NE has the property of being Pareto optimal.
Unfortunately there is no {\em a priori} justification to restricting
the space of quantum operators to those of with $\beta=0$.

With the full set of three parameter quantum strategies
every strategy has a counter strategy
that yields the opponent the maximum payoff of five,
while the player is left with the minimum of zero~\cite{benjamin00a}.
This result arises since for any $\hat{A} = \hat{U}(\theta,\alpha,\beta)$
there exists $\hat{B} = \hat{U}(\theta, \alpha, -\frac{\pi}{2} - \beta)$ such that
\begin{equation}
(\hat{A} \otimes \hat{I}) \frac{1}{\sqrt{2}} ( |00\rangle + i |11\rangle)
 = (\hat{I} \otimes \hat{B}) \frac{1}{\sqrt{2}} ( |00\rangle + i |11\rangle ) \;.
\end{equation}
That is, on the maximally entangled state
any unitary operation that Alice carries out on her qubit
is equivalent to a unitary operation that Bob carries out on his.
So for any strategy $\hat{U}(\theta, \alpha, \beta)$ chosen by Alice,
Bob has the counter  $\hat{D} \hat{U}(\theta, -\alpha, \frac{\pi}{2} - \beta)$,
essentially ``undoing'' Alice's move and then defecting.
Hence there is no equilibrium amongst pure quantum strategies.

We still have a (non-unique) NE amongst
mixed quantum strategies~\cite{eisert00}.
A mixed quantum strategy is the combination of two or more
pure quantum strategies using classical probabilities.
This is in contrast to a superposition of pure quantum strategies
which simply results in a different pure quantum strategy.
The idea is that Alice's strategy consists of choosing the pair of moves
\begin{equation}
\label{e-ne1}
\hat{A}_1 = \hat{C} = \left( \begin{array}{cc}
				1 & 0 \\
				0 & 1
		  	    \end{array} \right), \;\;
\hat{A}_2 = \left( \begin{array}{cc}
			i & 0 \\
			0 & -i
		  \end{array} \right)
\end{equation}
with equal probability,
while Bob counters with the corresponding pair of optimal answers
\begin{equation}
\label{e-ne2}
\hat{B}_1 = \hat{D} = \left( \begin{array}{cc}
				0 & i\\
				i & 0
		  	    \end{array} \right), \;\;
\hat{B}_2 = \left( \begin{array}{cc}
			0 & -1 \\
			1 & 0
		  \end{array} \right)
\end{equation}
with equal probability.
The combinations of strategies $\{A_i, B_j\}$ provide Bob
with the maximum payoff of five
and Alice with the minimum of zero when $i=j$,
while the payoffs are reversed when $i \ne j$.
The expectation value of the payoffs for each player is then
the average of $P_{CD}$ and $P_{DC}$, or 2.5.
There is a continuous set of NE of this type,
where Alice and Bob each play a pair of moves with equal probability, namely
\begin{eqnarray}
\hat{A}_1 = \hat{U}(\theta, \alpha, \beta) &,& \;\;
\hat{A}_2 = \hat{U}(\theta, \frac{\pi}{2} + \alpha, \frac{\pi}{2} + \beta) \;,
\\
\hat{B}_1 = \hat{U}(\pi - \theta, \frac{\pi}{2} + \beta, \alpha) &,& \;\;
\hat{B}_2 = \hat{U}(\pi - \theta, \pi + \beta, \frac{\pi}{2} + \alpha) \;.
\nonumber
\end{eqnarray}

If other values of the payoffs were chosen in Eq.~(\ref{e-pd}),
while still retaining the conditions for a classical prisoners' dilemma\footnote{A
prisoners' dilemma is characterized
by the payoffs for the row player being in the order
$P_{DC} > P_{CC} > P_{DD} > P_{CD}$.},
the average quantum NE payoff may be below (as is the case here) or above that
of mutual cooperation~\cite{benjamin00a}.
In the latter case the conflict between the NE and the Pareto optimal
outcome has disappeared,
while in the former we have at least
an improvement over the classical NE result of mutual defection.

In Ref.~\cite{du01a} a quantum prisoners' dilemma with Eisert {\em et al}'s scheme
was achieved on a two qubit nuclear magnetic resonance computer,
with various degrees of entanglement,
from a separable (i.e., classical) game to a maximally entangled quantum game.
Good agreement between theory and experiment was obtained.

The prescription provided by Eisert {\em et al} is a general one that can be
applied to any $2\times2$ game,
with the generalization to $2\times n$ games being
to use SU($n$) operators to represent the players' actions.

This method of quantization is not unique.
Another way of achieving similar results is simply to dispense with
the entanglement operators
and simply hypothesize various initial states,
an approach first used by Marinatto and Weber~\cite{marinatto00}
and since used by other authors~\cite{flitney02a,iqbal01b}.
The essential difference to Eisert's scheme is the absence of a disentangling operator.
Different games are obtained by assuming different initial states.
The classical game
(with quantum operators representing mixed classical strategies)
is obtained by selecting $|\psi\rangle = |00\rangle$,
while an initial state that is maximally entangled
gives rise to the maximum quantum effects.
In references~\cite{marinatto00,iqbal01b} the authors restrict
the available strategies to probabilistic mixtures of the identity and bit flip operators
forcing the players to play a mixed classical strategy.
The absence of the $\hat{J}^{\dagger}$ gate still leads to different results
than playing the game entirely classically.

Iqbal has considered ESS in quantum versions of both the prisoners' dilemma
and the battle of the sexes~\cite{iqbal01a}
and concluded that entanglement can be made to produce or eliminate ESSs
while retaining the same set of NE.
A classical ESS can easily be invaded by a mutant strategy that employs
quantum means and that can exploit entanglement.
Without the entanglement, the quantum mutants have no advantage.
In these models the replicator dynamic takes a ``quantum'' form~\cite{iqbal01}.

\subsection{Larger strategic spaces}
Since this initial work,
the field of quantum games has been extended to multiplayer games
and games with more than two pure classical strategies.
As situations become more complicated
there is more flexibility in the method of quantization.

Additional players are easily accommodated in the quantum game protocol
of Fig.~(\ref{f-game22}) by adding additional qubits to the initial state
and additional player operators,
as shown in Fig.~(\ref{f-gamen2}).
The entanglement operator of Eq.~({\ref{e-entangle})
creates a maximal entanglement between all the players' qubits.

\begin{figure}
\begin{picture}(180,120)(-80,0)
        \multiput(0,58)(0,30){2}{$|0\rangle$}
        \put(0,8){$|0\rangle$}
        \put(0,73){$\otimes$}
        \put(160,45){$|\psi_f\rangle$}

        \multiput(20,60)(40,0){4}{\line(1,0){20}}
        \multiput(20,90)(40,0){4}{\line(1,0){20}}
        \put(45,73){$\hat{J}$}
        \put(125,73){$\hat{J}^{\dagger}$}

	\multiput(40,50)(20,0){2}{\line(0,1){50}}
	\multiput(120,50)(20,0){2}{\line(0,1){50}}
	\multiput(40,100)(80,0){2}{\line(1,0){20}}

        \put(80,80){\framebox(20,20){$\hat{M}_1$}}
        \put(80,50){\framebox(20,20){$\hat{M}_2$}}
        \put(80,0){\framebox(20,20){$\hat{M}_N$}}

	\multiput(40,0)(80,0){2}{\line(1,0){20}}
	\multiput(40,0)(20,0){2}{\line(0,1){20}}
	\multiput(120,0)(20,0){2}{\line(0,1){20}}
        \multiput(20,10)(40,0){4}{\line(1,0){20}}

	\multiput(5,30)(85,0){2}{$\vdots$}
	\multiput(40,25)(0,8){3}{\line(0,1){4}}
	\multiput(60,25)(0,8){3}{\line(0,1){4}}
	\multiput(120,25)(0,8){3}{\line(0,1){4}}
	\multiput(140,25)(0,8){3}{\line(0,1){4}}
        \put(60,110){\vector(1,0){60}}
        \put(75,115){time}

\end{picture}
\caption{The flow of information in an N-person quantum game,
where $M_i$ is the move of the $i$th player
and $\hat{J}$ is an entangling gate.
From Ref.~\cite{benjamin00b}.}
\label{f-gamen2}
\end{figure}
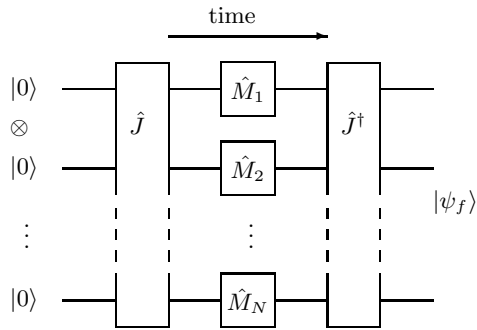

Benjamin and Hayden~\cite{benjamin00b} have examined three and four
player quantum games.
These are strategically richer than the two player ones.
For example,
it is possible to construct a prisoners' dilemma-like three handed game
that has a NE in pure quantum strategies
that is either better or worse than the classical one.

A game where entanglement can be exploited particularly effectively
is the minority game.
The players must select either zero or one.
If they select the least popular choice
they are rewarded.
No reward is given if the numbers are balanced.
Classically, the players can do no better than making a random selection
and the situation is not improved in the three player quantum version.
In the four player classical game half the time there is no minority,
so each player wins on average only one time in eight.
However, entanglement in the quantum version allows us to avoid this outcome
and provides a NE which rewards each player with probability one quarter,
twice the classical average~\cite{benjamin00b}.

Games with more than two classical pure strategies can be modeled
in ways similar to the strategically smaller games
by replacing the qubits representing the players' decisions
by, in general, an $n$-state quantum system
(or qunit) for the $n$-choice case.
The space of pure quantum strategies is expanded from SU(2) to SU($n$).

The game of rock-scissors-paper,
where the players have three choices,
has been examined by Iqbal and Toor~\cite{iqbal01b}.
However, to make the game amenable to analysis,
the authors do not allow the players the full range of unitary operations,
but rather restrict the strategies to mixtures of $\hat{I}$
and two operators that involve the interchange of a pair of states.
Entanglement still provides for an enrichment over the classical game.

There are three quantum versions of the game show situation known
as the Monty Hall problem~\cite{li01,flitney02a,dariano02}
in which players have a three way choice.
In the work by the present authors~\cite{flitney02a}
the game is modeled using suitable unitary operators,
with the participants having access to the full set of SU(3) operations.
In the quantum version
either player can exploit the entanglement
to their advantage if the other person employs only classical means
but if the full set of SU(3) operators are available to both players
we again have a situation where every strategy has a counter strategy.
It is still possible to find a (non-unique) NE amongst mixed quantum strategies.

Li {\em et al}~\cite{li01} permit one of the players to use an ancillary entangled
particle and to take measurements on this as part of their strategy.
Against a classical opponent,
this turns what was a biased game into a fair one.

The final version~\cite{dariano02} adopts a different protocol
for constructing a quantum game.
The author's dispense with the idea of unitary operators acting on qutrits
and instead have the participants directly selecting
vectors in a three dimensional Hilbert space.
In this variant a classical player can always be defeated by a quantum one.

Some of the mathematical methods of physics have attracted the attention
of economists and a new branch of economic mathematics has appeared, econophysics.
Recently, Polish theorists Piotrowski and S{\l}adkowski
have proposed a quantum-like approach to economics with its roots
in quantum game theory~\cite{piotr01a}.
Classical game theory is already extensively used by economists.
In the new quantum market games, transactions are described in terms
of projective operations acting on Hilbert spaces of strategies of traders.
A quantum strategy represents a superposition of trading actions
and can achieve outcomes not realizable by classical means~\cite{piotr02}.
Furthermore, quantum mechanics has features
that can be used to model aspects of market behavior.
For example, traders observe the actions of other players
and adjust their actions accordingly,
so there is non-commutativity of bidding~\cite{piotr01b},
maximal capital flow at a given price corresponds to entanglement between
buyers and sellers~\cite{piotr01a}, and so on.
There is speculation that markets cleared with quantum algorithms
will have increased efficiency~\cite{piotr01b}
and avoiding dramatic market reversals.

\subsection{Quantum Parrondo's games}
Parrondo's paradox, or Parrondo's games arise
when we have two games that are losing when played in isolation,
but when played in combination
form an overall winning game~\cite{harmer99a,mcclintock99}.
This necessarily involves some form of coupling between the games,
for example,
through the player's capital~\cite{harmer99b}
or via history dependent rules~\cite{parrondo00}.

There has been recent attempts to create quantum versions of Parrondo's games.
In references~\cite{ng01,flitney02b}
the history dependent game of Parrondo {\em et al}~\cite{parrondo00}
has been translated directly into the quantum sphere
by replacing the tossing of classical coins
by SU(2) operations on qubits.
There is coupling through the history dependent rules
with additional effects arising from interference
when a suitable superposition of states is chosen as the initial state.
Meyer and Blumer~\cite{meyer01} uses a quantum lattice gas automaton
to construct a Parrondo game
involving a single particle in an unbiased random walk between lattice sites.
Ratcheting in one direction is achieved by multiplication by a position
dependent phase factor
and the resulting quantum interference.
Lee and Johnson~\cite{lee02} have examine the relationship between
efficient quantum algorithms and Parrondo's games.
Here a random mixture of two algorithms produces a superior
result than either one alone.

\section{Discussion and open questions}
We have considered the basic protocol for simple quantum games
and given examples of the various possibilities
that have been discussed in the literature.
In general the quantum representation of a classical game is not unique
but all contain the classical game as a subset.
The full set of quantum operations can be represented
by trace-preserving, completely-positive maps.
The possibilities where those operations are not directly unitary,
such as the use of ancillas and the performance of measurements,
remain little explored.

Quantization of a game can lead
to either the appearance or disappearance of Nash equilibria.
In general the much enhanced strategic space available to the players
makes the quantum game more ``efficient'' than its classical counter part.
For example, the gap between the Pareto optimal outcome
and the NE in the prisoners' dilemma is reduced or eliminated,
and the average payoff in a multi-player minority game is increased,
when players are permitted to use (mixed) quantum strategies.
There are no equilibria in the space of pure quantum strategies
in an entangled, fair, $2 \times 2$ quantum game,
a result that is easily extended to $2 \times n$ games,
but general results for multiplayer games
or repeated games remain to be discovered.

Ng and Abbott~\cite{ng01} have posed the as yet unanswered question:
can coupling in a quantum Parrondo's game
be achieved through quantum entanglement alone?
We have already mentioned how the selection of different initial superpositions
can lead to different quantum games~\cite{marinatto00,iqbal01b,flitney02a}.
Is it possible to construct a quantum game where,
by choosing a suitable superposition for the initial state,
the resulting games can be individually losing
but the superposition of the results produces a positive payoff?
The coupling in this case would be through quantum interference,
arranged to minimize the amplitudes of the losing final states
and maximize those of the winning states.

The effect of a noisy input to
a quantum game has been examined by Johnson~\cite{johnson01}.
In this example a suitable level of noise enhances the payoff.
Decoherence can be simulated in a quantum game
by applying a controlled not gate to the desired qubit,
with the control bit being a random classical bit~\cite{benjamin00b}.
The effect of noise and decoherence
on a quantum game is another area that
invites further examination.

\section*{Acknowledgment}
This work was supported by GTECH Corporation Australia
with the assistance of the SA Lotteries Commission (Australia).

\end{document}